\documentclass[10pt]{article}
\usepackage{graphicx}

\def\Title#1{\begin{center} {\Large #1 } \end{center}}
\def\Author#1{\begin{center}{ \sc #1} \end{center}}
\def\Address#1{\begin{center}{ \it #1} \end{center}}

\newcommand\pubblock{\rightline{\begin{tabular}{l} Proceedings of the Fifth Annual LHCP\\ \pubnumber\\
         \pubdate  \end{tabular}}}

\newenvironment{Abstract}{\begin{quotation} \begin{center} 
             \large ABSTRACT \end{center}\bigskip 
      \begin{center}\begin{large}}{\end{large}\end{center} \end{quotation}}

\newenvironment{Presented}{\begin{quotation} \begin{center} 
             PRESENTED AT\end{center}\bigskip 
      \begin{center}\begin{large}}{\end{large}\end{center} \end{quotation}}

\def\beq{\begin{equation}}
\def\eeq#1{\label{#1}\end{equation}}
\def\eeqn{\end{equation}}

\def\beqa{\begin{eqnarray}}
\def\eeqa#1{\label{#1}\end{eqnarray}}
\def\eeqan{\end{eqnarray}}

\let\bar=\overbar

\def\Dslash{\not{\hbox{\kern-4pt $D$}}}
\def\dslash{\not{\hbox{\kern-2pt $\del$}}}

\def\msb{{\bar{\ssstyle M \kern -1pt S}}}

\usepackage{color}
\newcommand{\njets}{\ensuremath{N_{\mathrm{jets}}}}
\newcommand{\nb}{\ensuremath{N_{\mathrm{b-jets}}}}
\newcommand{\mttwol}{\ensuremath{M_{\mathrm{T2}}(\ell\ell)}}
\usepackage{multirow}

\newcommand\pubnumber{ CMS-CR-2017-222 }
\newcommand\pubdate{\today}

\def\affiliation{
on behalf of the CMS Collaboration, \\
Universidad de Oviedo, Spain}

\begin{document}

% large size for the first page
\large
\begin{titlepage}
\pubblock

%% Change the title, name, abstract
%% Title 
\vfill
\Title{  Searches for strong production of SUSY particles with two opposite-sign same-flavor leptons at CMS   }
\vfill

%  if you need to add the support use this, fill the \support definition above. 
%   \Author{ FIRSTNAME LASTNAME \support }
\Author{ SERGIO S\'ANCHEZ CRUZ,  }
\Address{\affiliation}
\vfill
\begin{Abstract}

A search is presented for physics beyond the standard model in events with two opposite-sign, same-flavor leptons, jets and missing transverse momentum in the final state. The search is performed in a dataset of 35.9 $\mathrm{fb}^{-1}$ of $\sqrt{s} = $ 13 TeV pp collisions recorded by the CMS experiment in the year 2016. The search targets models in which a colored particle is produced. Models are considered, in which a kinematic edge is observed in the dilepton invariant mass distribution and models in which a Z boson arises in the decay chain of the SUSY particles. Such searches have been performed in 8 TeV pp collisions as well as 13 TeV collisions. This version of the search adds additional event categories as well as improved background estimation procedures substantially increasing the sensitivity of the search. The results are interpreted in the context of simplified models of Supersymmetry. 

\end{Abstract}
\vfill

% DO NOT CHANGE 
\begin{Presented}
The Fifth Annual Conference\\
 on Large Hadron Collider Physics \\
Shanghai Jiao Tong University, Shanghai, China\\ 
May 15-20, 2017
\end{Presented}
\vfill
\end{titlepage}
\def\thefootnote{\fnsymbol{footnote}}
\setcounter{footnote}{0}
%

% normal size for the rest
\normalsize 

%% Your paper should be entered below. 

\section{Introduction}

This documents reports a search for new physics in events with two opposite-sign,
same-flavor leptons, jets and missing transverse momentum in pp collisions recorded
by the CMS experiment\cite{cms}. These searches have been performed by both the CMS\cite{edge_cms1,edge_cms2} and ATLAS Collaborations\cite{edge_atlas1,edge_atlas2}, and are motivated by several models of supersymmetry (SUSY) that 
involve the production of such lepton pairs in either the decay of an on-shell Z boson
produced in the decay chain of SUSY particles or in the sequential two-body decays
of a neutralino. 

These two topologies correspond to different experimental signatures:
while the former would be observed as an excess in events compatible with a Z boson, the
latter would yield to a edge-shape feature in the dilepton invariant mass, $m_{\ell\ell}$, distribution. 

\begin{figure}[htb]
\centering
\includegraphics[width=0.4\textwidth]{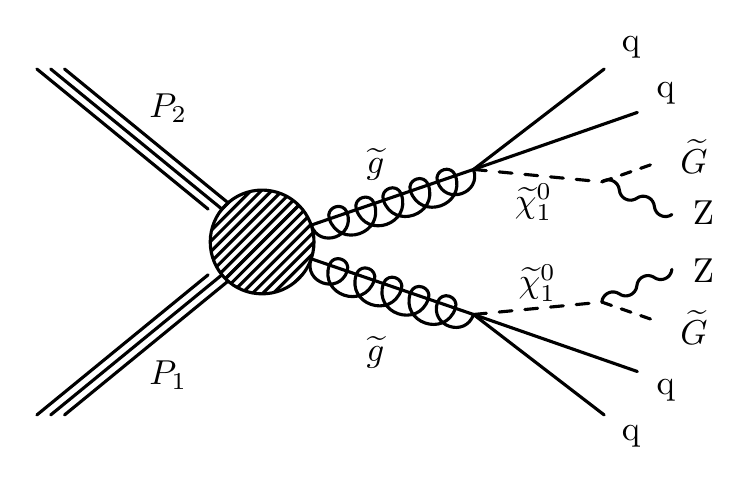}
\includegraphics[width=0.4\textwidth]{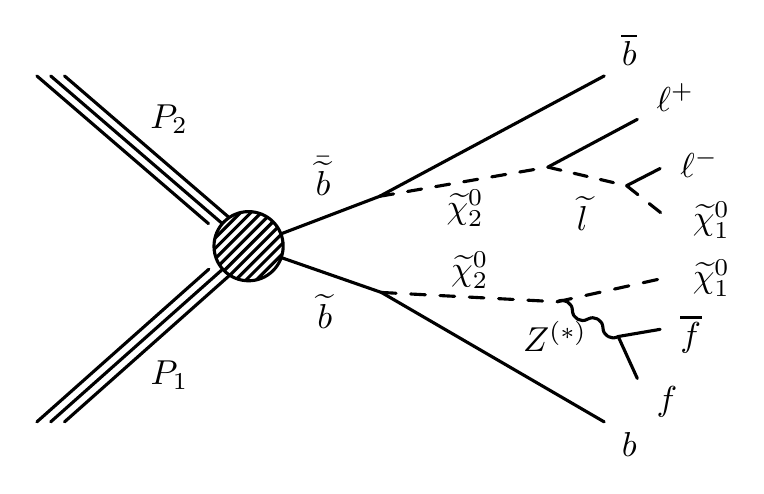}
\caption{ Diagrams of the simplified SUSY models used for the interpretation of the 
results of the analysis. Diagram in the left shows the GMSB scenario and the diagram on the right represents the slepton-edge model.  }
\label{fig:feynman}
\end{figure}

Figure~\ref{fig:feynman} shows the diagrams of the simplified models used for 
the interpretation of the results of this analysis. In such models the production and decay 
of a few SUSY particles is taken into account, assuming the rest of the SUSY particles
are beyond the experiment scope. 
The first simplified model, which represents a Gauge Mediated Supersymmetry breaking model and is referred to as the GMSB scenario, involves the production of two gluinos decaying into quarks and a neutralino, that  can further decay into an on-shell Z boson and a 
massless gravitino. In the other simplified model considered, the slepton-edge scenario, two sbottoms are produced, decaying 
into $b$-quarks and  neutralinos, that further decay into off-shell Z bosons and the LSP. 

\section{Search strategy and background estimation}

Events with two opposite-sign same-flavor leptons, two jets and missing transverse momentum 
($E_T^{miss}$) greater than 100 GeV are considered in this analysis.
On top of this common baseline selection, the spectrum of the dilepton invariant mass
is splitted into an on-Z part, sensitive to the first kind of models described in the 
introduction, and an off-Z part, sensitive to the other kind. 

The background estimation methods, described in this section, are also  
applied in the same way to both parts of the analysis.

The main background contribution to many of the search regions that will
be described is due to so-called flavor-symmetric processes, i.e. those processes
that yield a same-flavor or an opposite-flavor dilepton pair with the 
same probability. This class of backgrounds is dominated by $t\bar{t}$ production
and is estimated in a fully data-driven fashion. In order to do so, the opposite-flavor
channel is used as a side-band, and the obtained estimate is corrected by the 
different electron and muon trigger and identification efficiencies. 

Even if they do not produce invisible particles in the final state, 
Drell-Yan events can contain instrumental missing transverse momentum and enter the defined baseline region. This instrumental missing transverse 
momentum is usually driven by the resolution in the measurement of the 
jet energy. In order to estimate this contribution, the  
$E_T^{miss}$ distribution  is determined in a dedicated $\gamma$ + jets
data sample, in which the instrumental $E_T^{miss}$ is also driven by the 
jet resolution.

Remaining contributions from backgrounds with a Z boson and genuine $E_T^{miss}$
(WZ, ZZ, $t\bar{t}Z$) are estimated using dedicated Monte Carlo samples 
that are validated in  background enriched control regions.

\section{Search regions}
In order to be sensitive to a wide range of SUSY particle masses, two sets of 
orthogonal signal regions are defined on top of the on-Z and off-Z search regions. 

Events in the on-Z category, with $|m_{\ell\ell} - 91$ GeV$|$  $<$ 5 GeV are classified
according to the hadronic activity, the number of $b$-jets and
$E_T^{miss}$. 

Events in the off-Z category are further required to have $E_T^{miss} > $ 150 GeV
and the kinematic observable $m_{T2}$\cite{mt2} to be greater than 80 GeV. Events are then further classified according to the 
$m_{\ell\ell}$  of the reconstructed dilepton system, in order to target 
different locations of the signal kinematic edge.

Additionally, events are classified as $t\bar{t}$-like and
non-$t\bar{t}$-like according to a Negative Log Likelihood (NLL) discriminant, which is a likelihood
discriminant  designed to reject events due flavor-symmetric processes, 
the largest contribution in the off-Z regions. The observables used for the 
likelihood discriminant are: the $p_{T}$ of the dilepton system, $|\Delta(\phi)|$ between
the leptons, $E_T^{miss}$ and an observable called $\sum m_{lb}$. The latter is 
calculated by taking all possible combinations of leptons and $b$-jets, 
finding the combination with the minimum mass, $m_{lb}$. If no $b$-jets are found, the combinations
of leptons and light jets are considered. Then the process is repeated with the remaining lepton and
$b$-jets.  $\sum m_{lb}$ is defined as the sum of those masses. 

In order to define the NLL discriminant, a parametric model for the probability distribution functions (pdfs) 
of each of the four observables is assumed. The particular pdf is then extracted
by fitting the parametric model to data in the opposite-flavor side-band, 
which is $t\bar{t}$ enriched. Then, for each event the NLL discriminant can be
calculated as 

\[
\mathrm{NLL} = -\sum_i \log f_i(x_i),
\]
where $x_i$ are the four observables taken into account and $f_i$ are their fitted 
pdfs. The NLL coincides with the opposite of the logarithm of the likelihood for an event been
produced in a $t\bar{t}$  process in the approximation in which the $x_i$ variables 
are independent. These  four variables were verified to be only mildly correlated and
the NLL has been observed to be a powerful discriminant to a wide range of
parameter sets in the slepton-edge model. 

Based on the value of the NLL discriminant, a $t\bar{t}$-like and a 
non-$t\bar{t}$-like region are defined. 

The complete list of search regions is shown in tab.~\ref{tab:selections_signalRegions}.

\begin{table}[htb]
\begin{center}
\caption{\label{tab:selections_signalRegions} Summary of all search regions. $H_T$ is defined as the scalar sum of the transverse momenta of the selected jet.}
\footnotesize
\begin{tabular}{l|l|l|l|l|l}
\hline                                          
\multicolumn{6}{c}{ Strong-production on-Z ($86 < m_{\ell\ell} < 96$ GeV) search regions  }  \\
\hline                                         
Region & \njets & \nb & $H_T$ [GeV] & \mttwol [GeV] & $E_T^{miss}$ binning [GeV] \\
\hline                                         
SRA b veto & 2--3 & $= 0$ & $> 500$ & $> 80$ & 100--150, 150--250, $>250$ \\
SRB b veto & 4--5 & $= 0$ & $> 500$ & $> 80$ & 100--150, 150--250, $>250$ \\
SRC b veto & $\geq6$ & $= 0$ & - & $> 80$ & 100--150, $>150$  \\
SRA b tag  & 2--3 & $\geq 1$ & $> 200$ & $> 100$ & 100--150, 150--250, $>250$ \\
SRB b tag  & 4--5 & $\geq 1$ & $> 200$ & $> 100$ & 100--150, 150--250, $>250$ \\
SRC b tag  & $\geq6$ & $\geq 1$ & - & $> 100$ & 100--150, $>150$  \\

\hline
\multicolumn{6}{c}{Off-Z search regions}  \\
\hline
Region & \njets & $E_T^{miss}$ [GeV] & \mttwol [GeV] & NLL & $m_{\ell\ell}$  binning [GeV] \\
\hline 

  \rule{0pt}{2ex} \multirow{2}{*}{$t\bar{t}$-like} & \multirow{2}{*}{$\geq 2$} & \multirow{2}{*}{$> 150$} & \multirow{2}{*}{$> 80$} & \multirow{2}{*}{$< 21$} & 20--60, 60--86, 96--150, 150--200, \\
   & & & & & 200--300, 300--400, $>400$ \\

%  \multirow{2}{*}{not-\ttbar-like} & \multirow{2}{*}{$\geq 2$} & \multirow{2}{*}{$> 150$} & \multirow{2}{*}{$> 80$} & \multirow{2}{*}{$> 21$} & 20--60, 60--86, 96--150, 150--200, \\
%   & & & & & 200--300, 300--400, $>400$ \\
  \rule{0pt}{2ex} not-$t\bar{t}$-like & $\geq 2$ & $> 150$ & $> 80$ & $> 21$ & same as $t\bar{t}$-like \\
\hline                                          
\end{tabular}
\end{center}
\end{table}

\section{Results and interpretation}

\begin{figure}[htb]
\centering
\includegraphics[width=0.4\textwidth]{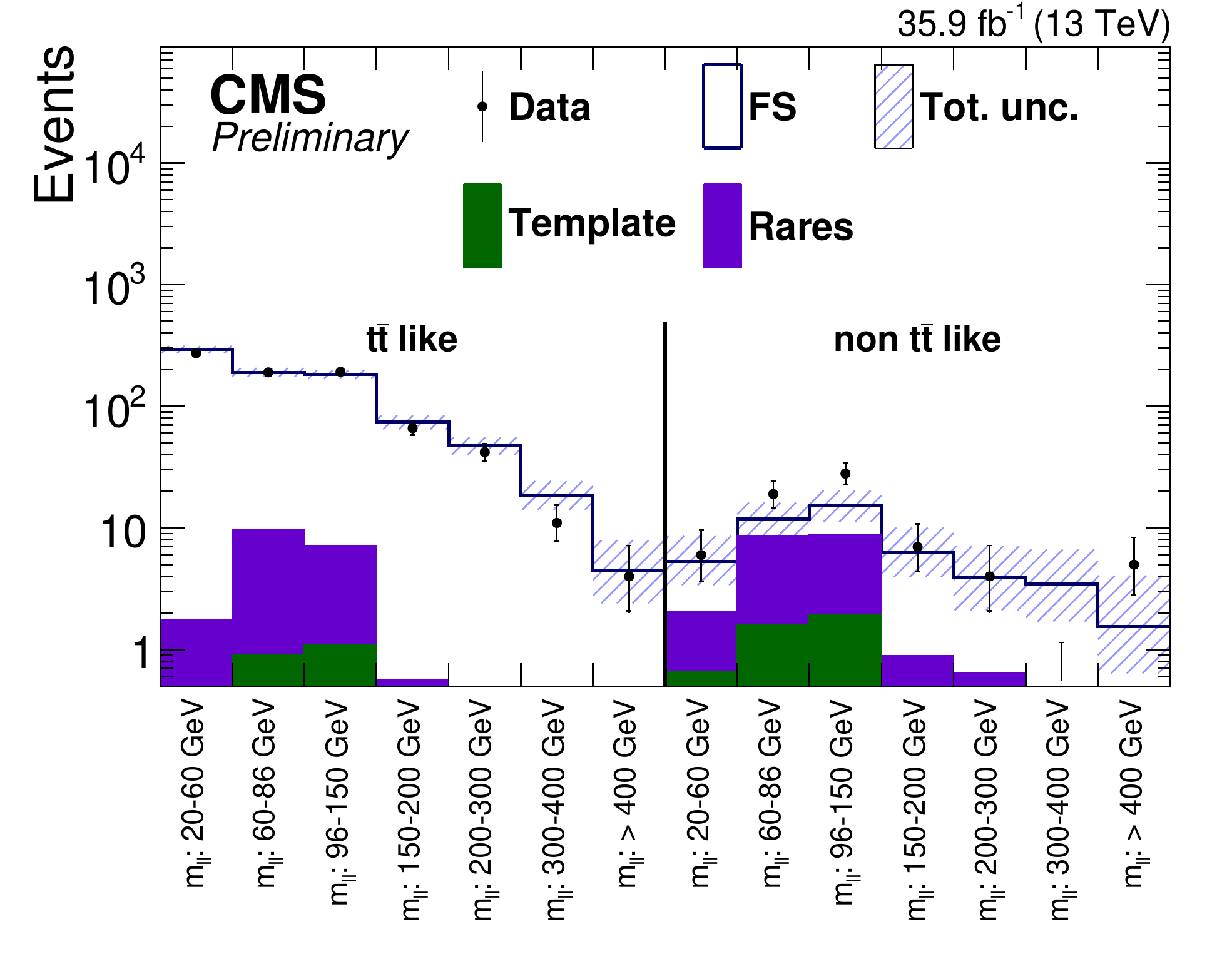}
\caption{ Results of the off-Z part of the search. For each search region, the number
of observed events (black dots) is compared to the total number of expected events 
(blue line).   }
\label{fig:results_off}
\end{figure}

\begin{figure}[htb]
\centering
\includegraphics[width=0.23\textwidth]{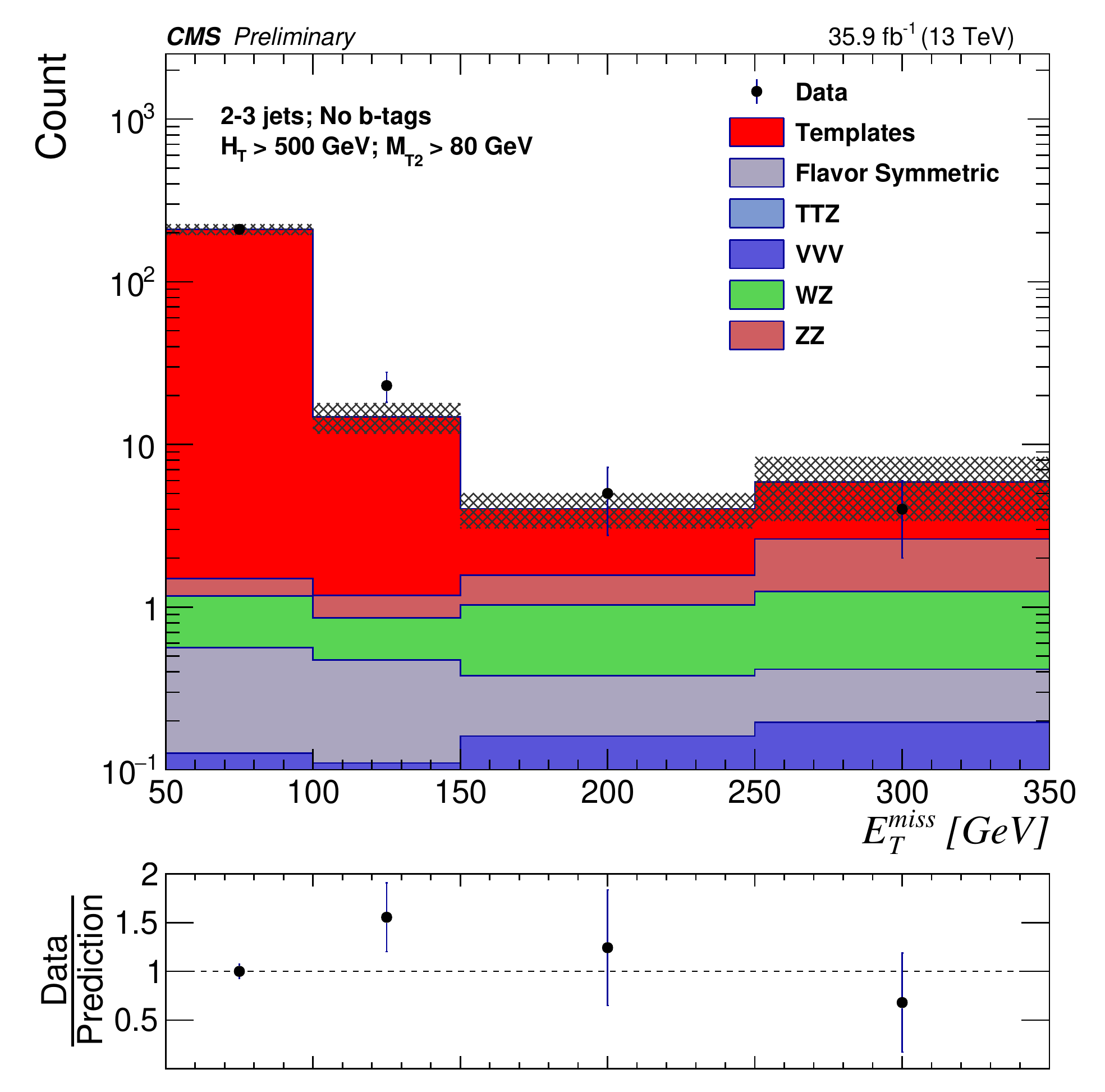}
\includegraphics[width=0.23\textwidth]{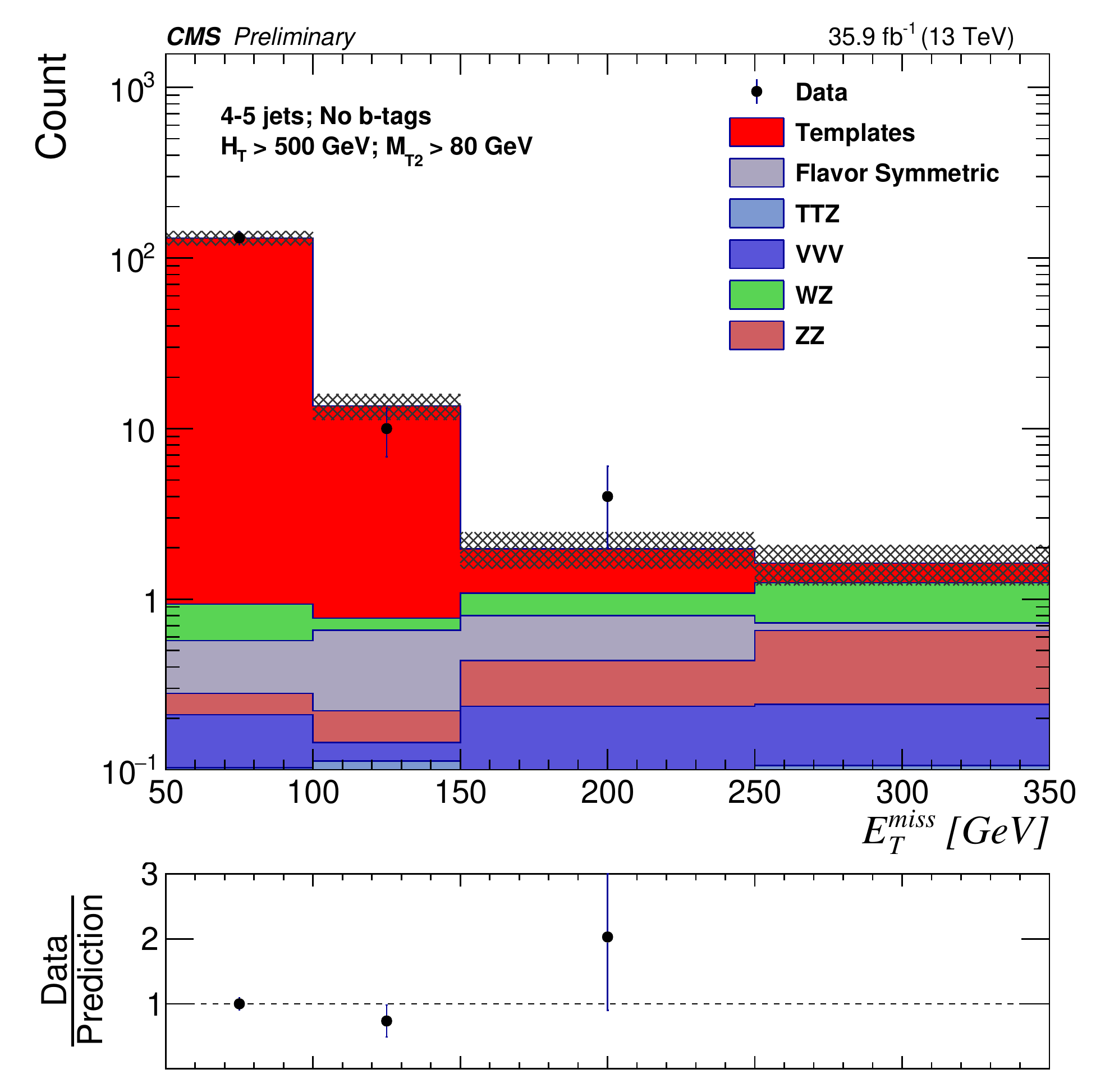}
\includegraphics[width=0.23\textwidth]{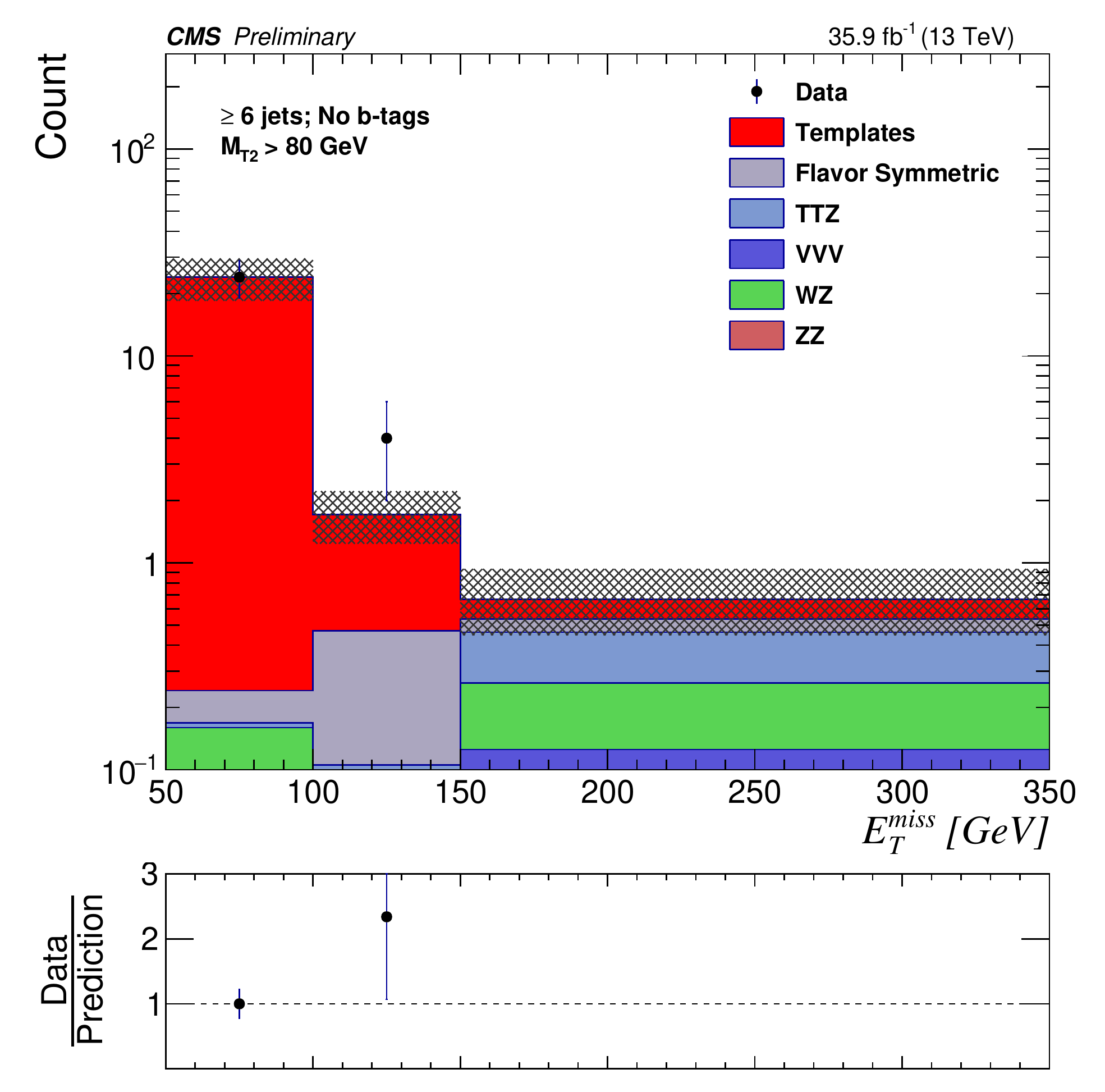}
                         
\includegraphics[width=0.23\textwidth]{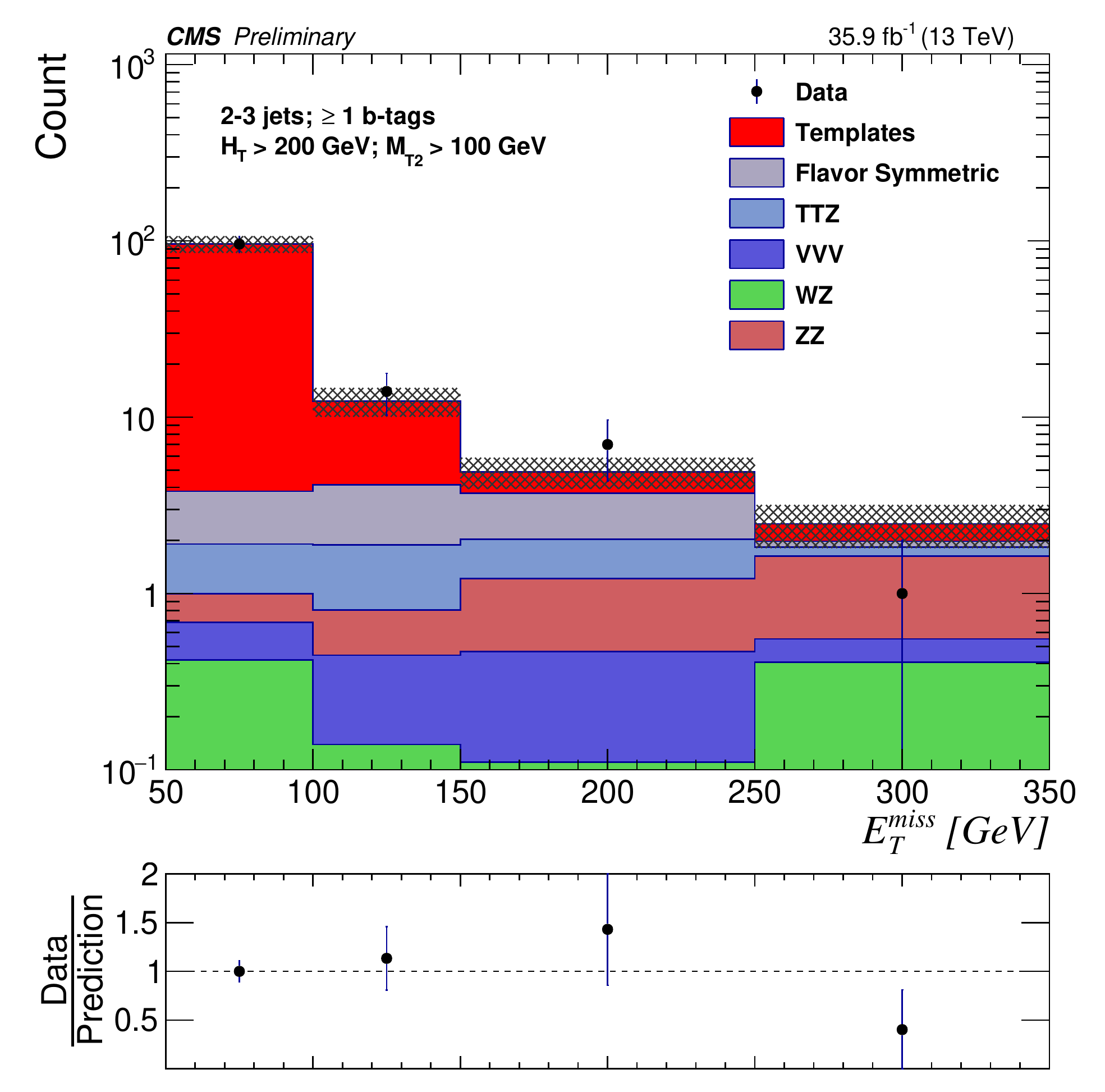}
\includegraphics[width=0.23\textwidth]{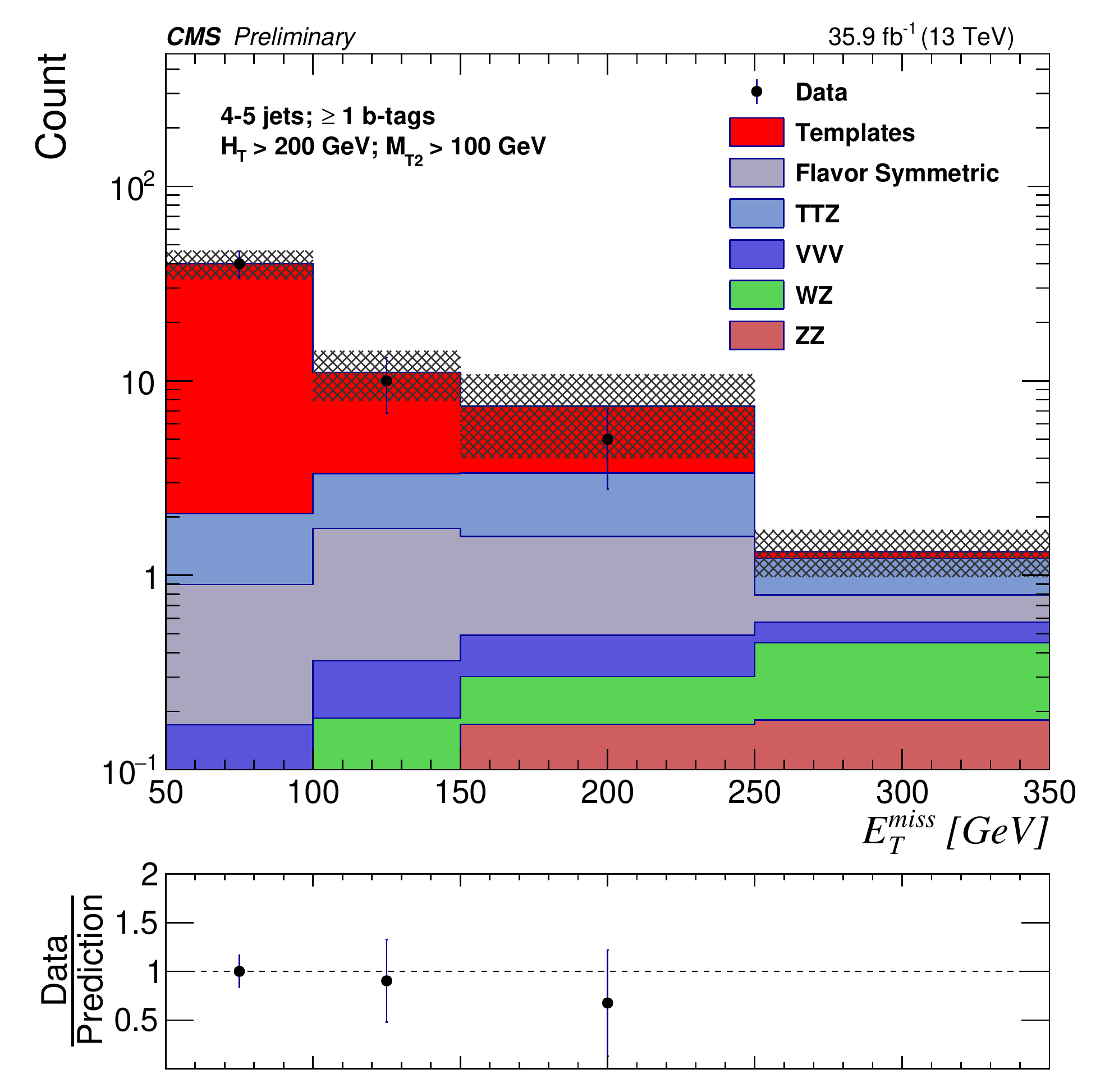}
OA\includegraphics[width=0.23\textwidth]{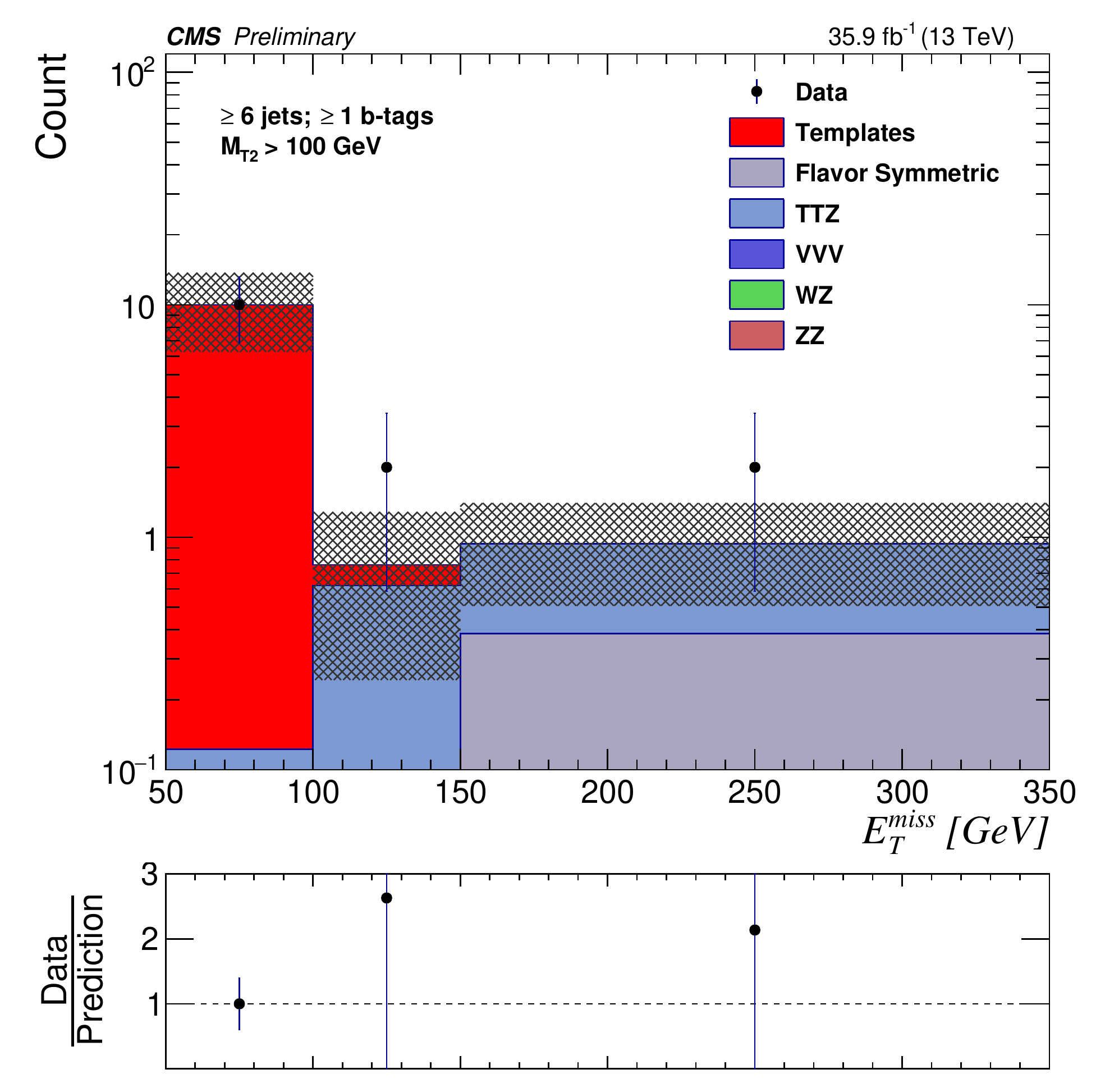}
\caption{ Results of the on-Z part of the search. Figures show the $E_T^{miss}$ 
distributions in events in search regions A (left), B (center) and C (right)
without $b$-jets (top) and with $b$-jets (bottom).}

\label{fig:results_on}

\includegraphics[width=0.28\textwidth]{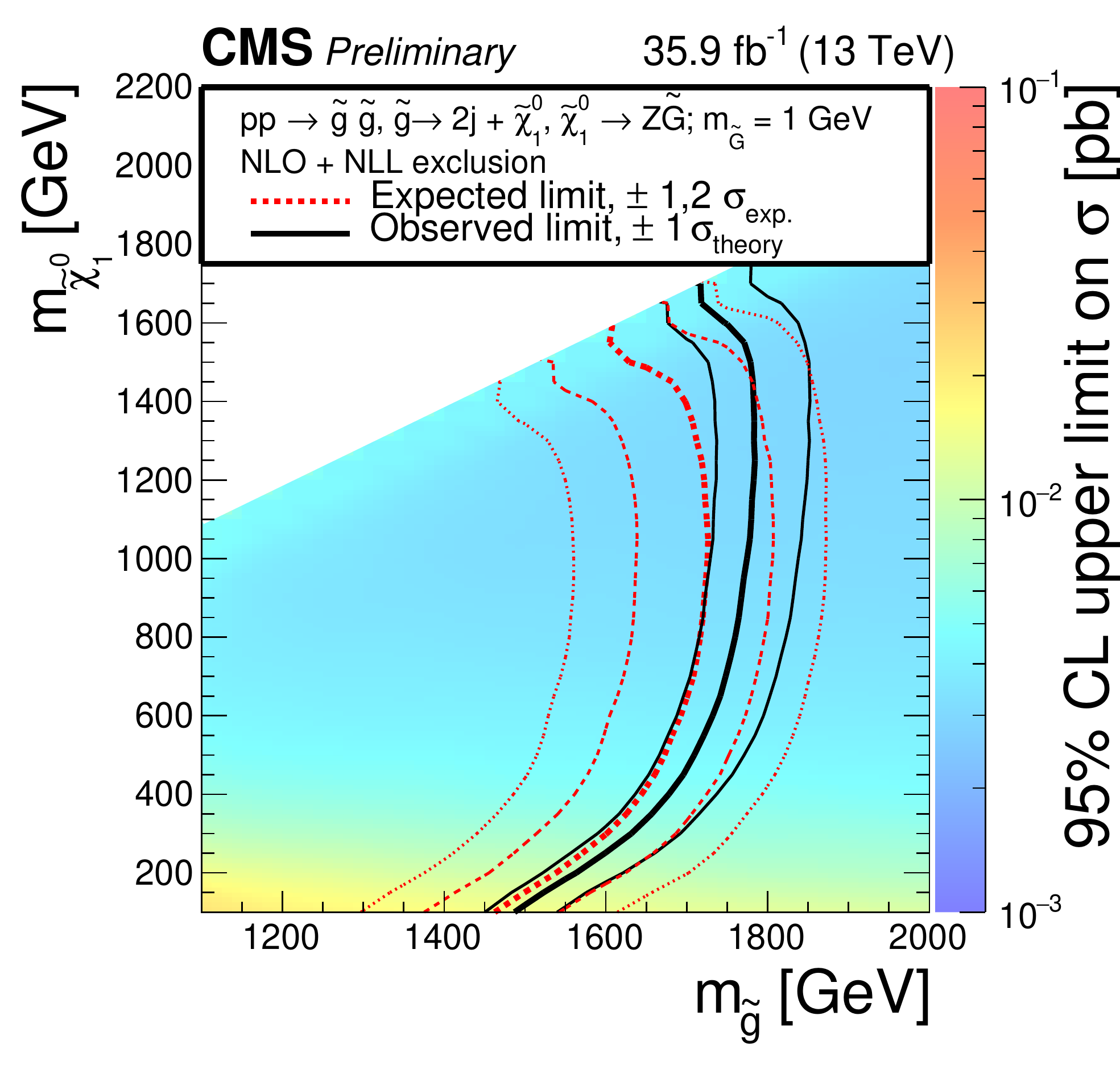}
\includegraphics[width=0.3\textwidth]{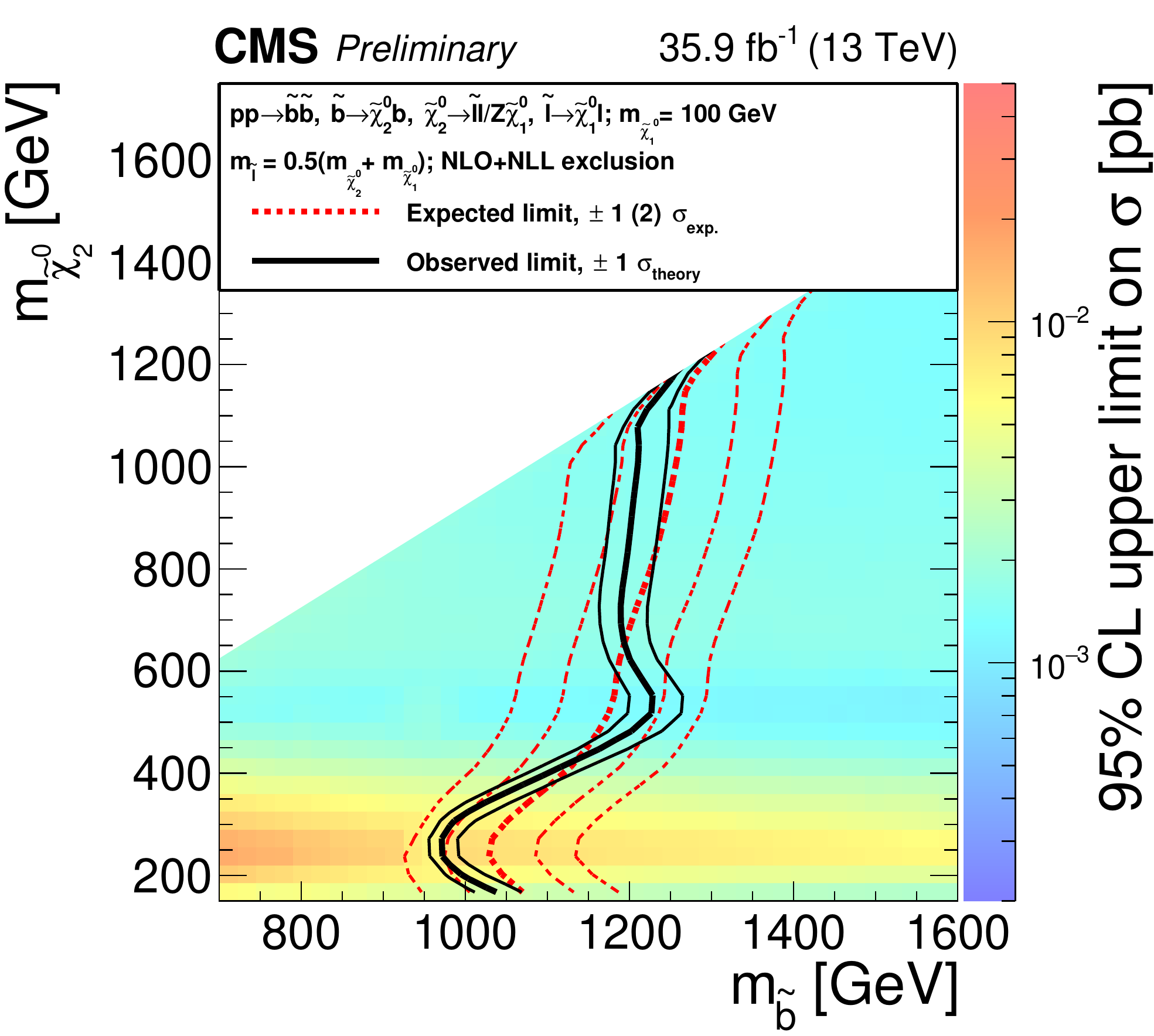}
\caption{The upper limit on the production cross-section are shown of the GMSB scenario as a function of the gluino and neutralino masses (left)
	and the slepton-edge scenario as a function of the sbottom and neutralino masses
	(right). The upper limit is shown in the color scale, while the red and black
	lines shown the expected and observed exclusion limits  on the masses based on a reference SUSY cross section. }
\label{fig:limits}

\end{figure}

Figure~\ref{fig:results_off} shows the observed and expected number of events 
in the off-Z search regions, and in Fig.~\ref{fig:results_on} the same is shown
for the on-Z regions. All the results are compatible with SM expectations.

No sign of Beyond Standard Model Physics has been observed.
Using these results, upper limits on cross sections and lower limits on masses can be set, assuming 100\% branching fractions for the models under consideration.
These limits are shown in Fig.~\ref{fig:limits}. 
In the GMSB senario, we exclude gluino masses up to 1500--1700 
GeV depending on the neutralino mass. In the slepton-edge
scenario, we exclude sbottom masses up to 1000--1200 GeV  depending on
the neutralino mass.

%\begin{figure}[htb]
%\centering
%\end{figure}

\section{Conclusions}

A search for new physics has been performed in events with two opposite-sign same-flavor 
lepton, jets and missing transverse momentum. The analysis targets two 
different kind of topologies, one in which the signal would be observed as an excess of 
events compatible with a Z boson, while in the other an edge-like feature would be observed
in the di-lepton mass distribution. Those topologies are motivated by simplified SUSY 
models.

The results obtained are compatible with Standard Model expectations, allowing us to exclude
gluino masses below 1500-1700 GeV and sbottom masses below 1000 to 1200 GeV in these simplified 
models.


\begin{thebibliography}{99}

%%
%%  bibliographic items can be constructed using the LaTeX format in SPIRES:
%%    see    http://www.slac.stanford.edu/spires/hep/latex.html
%%  SPIRES will also supply the CITATION line information; please include it.
%%
\bibitem{cms}
CMS Collaboration, JINST 3 S08004 (2008)

\bibitem{edge_cms1}
CMS Collaboration, Search for physics beyond the standard model in events with two  leptons, jets, and missing transverse momentum in pp collisions at $\sqrt{s} =$ 8 TeV, JHEP 04 (2015) 124

\bibitem{edge_cms2}
CMS Collaboration, Search for new physics in final states with two opposite-sign, same-flavor leptons, jets, and missing transverse momentum in pp collisions at $\sqrt{s}$=13 TeV. J. High Energy Phys. 12 (2016) 013


\bibitem{edge_atlas1}
ATLAS Collaboration, Search for supersymmetry in events containing a same-flavour opposite-sign dilepton pair, jets, and large missing transverse momentum in $\sqrt{s} = 8$ TeV pp collisions with the ATLAS detector. Eur. Phys. J. C 75 (2015) 318

\bibitem{edge_atlas2}
ATLAS Collaboration, Search for new phenomena in events containing a same-flavour opposite-sign dilepton pair, jets, and large missing transverse momentum in $\sqrt{s}$=13 TeV pp collisions with the ATLAS detector. Eur. Phys. J C 77 (2017) 144

\bibitem{mt2}
C. G. Lester and D. J. Summers, Measuring masses of semiinvisibly decaying particles pair produced at hadron colliders, Phys. Lett. B 463 (1999) 99

\bibitem{Aad:2012tfa} 
CMS Collaboration,  Search for new physics in final states with two opposite-sign, same-flavor
lepton, jets, and missing transverse momentum in pp collisions at $\sqrt{s}=$ 13 TeV. 
CMS-PAS-SUS-16-034 



\end{thebibliography}
\end{document}